\documentstyle[preprint,aps,epsf]{revtex}
\begin{document}
\draft
\title{Mesoscopic superpositions of states - approach to classicality and
diagonalization in coherent state basis}
\author{G. S. Agarwal\footnote{also at Jawaharlal Nehru Centre for Advanced
Scientific Research, Bangalore, India}}
\address{Physical Research Laboratory, Navrangpura, Ahmedabad-380009, India}
\maketitle
\begin{abstract}
I consider the interaction of a superposition of mesoscopic coherent states and
its approach to a mixed state as a result of a suitably controlled environment.
I show how the presence of a gain medium in a cavity can lead to diagonalization
in coherent state basis in contrast to the standard model of decoherence. I
further show how the new model of decoherence can lead to the generation of
$s$ ordered quasi distributions.
\end{abstract}
\newpage
	Mesoscopic superpositions of coherent states have been the subject of extensive
studies \cite{{one},{two},{three},{four},{five},{six},{seven}} because of their unusual interference characteristics and because
of their relevance to the quantum measurement problem. These states are also
known to be extremely sensitive to environmental interactions. The interference
terms disappear fast and a kind of diagonalization takes place \cite{{one},{four},{five}}. The
diagonalization is itself sensitive to the nature of the bath or the nature of
the interaction with environment. If the initial state is a superposition of
coherent states then ideally one would like to have a situation where the
interaction with the bath produces a mixed state involving the two coherent
states \cite{six}. There are however difficulties as the bath itself has certain
intrinsic properties \cite{eight}  which must be satisfied and these intrinsic properties
determine the dynamical characteristics of the subsystem. In this paper we
examine the question - how a manipulation of the bath could possibly produce a
diagonalization in coherent state basis.
	
	We note that the subject of the manipulation of the bath has also attracted
quite a bit of attention. Raimond {\it et al.} \cite{five}{ demonstrated how the coupling of a
high Q cavity containing the cat state to another resonator leads to the revival
of coherence. Several authors \cite{nine} have shown how the feedback and other mechanisms could
stabilize effects of decoherence. Poyatos {\it et al.} \cite{ten} demonstrated the engineering of the bath in
the context of laser cooled trapped ions. There are other models of decoherence
where the nonlinearities could give rise to coherence characteristics and could
indeed produce new types of states\cite{eleven}. Furthermore there exists the
possibility \cite{twelve} of achieving a control of the drift and diffusion terms in the
dissipative dynamics by external electromagnetic field. The external fields make
the environment {\it nonthermal} leading even to the possibility of making the
{\it drift term vanish} and {\it diffusion term rather small}. There are several
physical realizations of such pumped or nonthermal environment \cite{thirteen}.

	In this paper we consider the interaction of the field mode in a
mesoscopic superposition state with a bath which consists of a gain medium in
addition to the usual absorber. By choosing the gain appropriately we get purely
diffusive motion of the field mode. This motion leads to diagonalization in
coherent state basis though each coherent peak broadens due to diffusion. We
also demonstrate how the time evolution under purely diffusive motion leads to the
generation of the s-ordered quasi-distributions associated with the state of the
field.

	We start from a cat state, say, even or odd cat state for a bosonic
system 
\begin{equation}
\mid\psi\rangle={\cal N}_{\pm}(\mid\beta\rangle\pm\mid-\beta\rangle),	
\end{equation}
where the normalization constant is given by
\begin{equation}
{\cal N}_\pm^{-2}\equiv 2(1\pm\exp(-2\mid\beta\mid^{2})).		
\end{equation}
The bosonic mode may, for example, represent a field mode in a cavity or the
center of mass motion of an ion in a trap. 
The Wigner function $\Phi(\alpha,\alpha^*)$ for the state (1) is
\begin{eqnarray}
\Phi(\alpha,\alpha^*) &=&
\frac{2N_{\pm}^2}{\pi}(\exp\{-2\mid\alpha-\beta\mid^2\}
+\exp\{-2\mid\alpha+\beta\mid^2\}\nonumber\\ 
&\pm& 2\exp(-2\mid\alpha\mid^2)\cos(4\beta y));
~~~~\alpha=x+iy,~~\beta={\rm real}. 
\end{eqnarray}
The Wigner function thus consists of two Gaussians centered at $\alpha=\pm\beta$
with an interference term centered at the origin $\alpha=0$. The period of
oscillation depends on $\beta$. The interaction with the environment is
generally described by the density matrix equation [8] for the bosonic mode $a$
\begin{equation}
\frac{\partial\rho}{\partial t}=-\kappa(a^\dag a\rho-2a\rho a^\dag 
+\rho a^\dag a),
\end{equation}
where $2\kappa$ will be the rate of dissipation. The Wigner function at time $t$ will be given by
\begin{eqnarray}
\Phi(\alpha,\alpha^*,t)=
\frac{2{\cal N}_\pm^2}{\pi}(\exp(-2\mid\alpha-\beta e^{-\kappa t}\mid^2)
+\exp(-2\mid\alpha + \beta e^{-\kappa t}\mid^2) \nonumber \\
\pm 2\exp(-2\mid\alpha\mid^2)\exp(-2\beta^2(1-e^{-2\kappa t})) 
\cos(4\beta ye^{-\kappa t})).
\end{eqnarray}
We note that as a result of interaction with the environment the two Gaussians
move towards each other eventually merging into one Gaussian. The amplitude of
the oscillatory term goes down by a factor $\exp (-2\beta^2(1-e^{-2\kappa t}))$ and 
the period of oscillation increases by $e^{\kappa t}$. For $\kappa t\gg 1$, Eq. (5) goes over to 
\begin{equation}
\Phi \rightarrow \frac{2}{\pi} e^{-2\mid\alpha\mid^2}.
\end{equation}
For completeness we show this evolution in Fig.1 for different values of 
$\kappa t$.

	Intuitively, the emergence of classical behavior \cite{{six},{fourteen}} on interaction
with the environment would require a different behavior - we would expect to see a
double Gaussian structure with the missing oscillatory behavior. A natural
question arises what model of environment could achieve that. One natural
possibility is to consider a situation so that the exponentially damped factors
can be removed. For example, one could think of inserting a gain media in the
context of cavity problems. The gain can be chosen so as to compensate the loss.
Thus one might be able to keep the double Gaussian structure. However, any gain also
introduces some noise. We thus examine in detail the consequences of both
gain and loss on the dynamics of a mesoscopic superimposition of states. Let
$2\Gamma$ be the gain of the gain medium. Then Eq.(3) is modified to 
\begin{equation}
\dot{\rho}=-\kappa(a^{\dag}a\rho-2a\rho a^{\dag}+\rho a^{\dagger}a)-\Gamma
(aa^{\dag}\rho - 2a^{\dag}\rho a + \rho aa^{\dag}).
\end{equation}
The Wigner function obeys the equation of motion
\begin{equation}
\frac{\partial\Phi}{\partial t}=
(\kappa-\Gamma)\frac{\partial}{\partial\alpha}(\alpha\Phi)
+\frac{\kappa+\Gamma}{2}
\frac{\partial^2\Phi}{\partial\alpha\partial\alpha^*} + c.c..
\end{equation}
On writing $\alpha=x+iy,$ we get
\begin{equation}
\frac{\partial\Phi}{\partial t}=
(\kappa-\Gamma)\frac{\partial}{\partial x}(x\Phi)+
(\kappa-\Gamma)\frac{\partial}{\partial y}(y\Phi)+
\left( \frac{\kappa+\Gamma}{4}\right)
\left(\frac{\partial^2}{\partial x^2}+\frac{\partial^2}{\partial y^2}\right)\Phi.
\end{equation}
We now have two parameters $\kappa$ and $\Gamma$ which could be manipulated
independently to produce the desired result.

Now the drift and diffusion coefficients are respectively equal
to $(\kappa-\Gamma)$ and $(\kappa+\Gamma)/4$. We now have the possibility
of making drift {\it vanish} by choosing $\kappa=\Gamma$ leading to
\begin{equation}
\frac{\partial\Phi}{\partial t}=
2\kappa\frac{\partial^{2}\Phi}{\partial\alpha\partial\alpha^*}.
\end{equation}
The general solution of (10) can be expressed as 
\begin{equation}
\Phi(\alpha,\alpha^{*},t)\equiv\frac{1}{\pi\delta}\int\exp(-\mid\alpha-\alpha_{0}\mid^{2}/\delta)
\Phi(\alpha_{0},\alpha_{0}^{*},0)d^{2}\alpha_{0}; ~~   \delta=2\kappa t.
\end{equation}
On substituting (3) in (11) and on using the identity
\begin{equation}
\int\frac{d^2z}{\pi}\exp(\alpha z+\beta z^{*}-\gamma\mid z\mid^2)=
\frac{1}{\gamma}\exp\left(\frac{\alpha\beta}{\gamma}\right),
\end{equation}
we get
\begin{eqnarray}
\Phi(\alpha,\alpha^*,t) &=& \frac{2{\cal N}^2}{\pi(1+2\delta)}
(\exp\{-\frac{2}{(2\delta+1)}\mid\alpha -\beta\mid^2\}
+\exp\{-\frac{2}{(2\delta+1)}\mid\alpha+\beta\mid^2\} \nonumber \\
& &\pm 2\exp\{-\frac{2\mid\alpha\mid^2}{(1+2\delta)}
-\frac{4\beta^2\delta}{1+2\delta}\}\cos(\frac{4\beta y}{1+2\delta})
).
\end{eqnarray}
This result should be compared with the standard model Eq.(5) of decoherence.

	Thus for the interaction of a field mode in a Cat state with the new {\it environmental
conditions}, each component in the Wigner function remains {\it located} at the
{\it original position} as there is {\it no drift} in the model. However, each component undergoes
diffusion. For the usual model of decoherence there is no diffusion although the
mean position quickly drifts towards origin. The period of oscillation of the
interference term increases. The amplitude of oscillation also decreases. For
larger $\delta$ and for $\beta^2 >\delta+\frac{1}{2}$, the oscillatory (interference)
term disappears leading to 
\begin{equation}
\Phi(\alpha,\alpha^*, t)\approx\frac{2{\cal N}_{\pm}^2}{\pi(1+2\delta)}
\left( \exp \left( -\frac{2}{(2\delta +1)}\mid \alpha -\beta\mid^2\right) 
+\beta\rightarrow -\beta\right).
\end{equation}

We thus achieve diagonalization in coherent state basis - the decoherence to a mixed state which is a superposition of two Gaussians at $\pm\beta$. 
This is what we had set out to achieve. We show in Fig. 1 the effects of
decoherence on the Wigner function of the field mode interacting with this new
model of the environment. These results should be compared with the ones for the standard model of
decoherence. There are obviously important differences in the dynamics of a Cat
state interacting with different types of environment. 

We next present some very general results on various quasi distributions like
the P-function, the Q-function and the Wigner function. We discuss the parameter regime in which the {\it
nonclassical} characteristics of the original state start {\it disappearing}. For this purpose we examine the equation of
motion for the characteristic function $\langle\exp (\gamma a^\dag - \gamma^*a)
\rangle$ which is the Fourier transform of $\Phi$. Clearly the characteristic function
obeys the equation
\begin{equation}
\frac{\partial}{\partial t} \langle \exp (\gamma a^{\dag}
-\gamma^*a)\rangle =
-2\kappa\mid\gamma\mid^2
\langle \exp(\gamma a^\dag -\gamma^*a)\rangle,
\end{equation}
and hence
\begin{equation}
\langle\exp (\gamma a^\dag (t) - \gamma^* a(t)) \rangle =
\exp (-2\kappa\mid\gamma\mid^2 t)
\langle \exp (\gamma a^{\dag}-\gamma^* a)\rangle,
\end{equation}
which on using the disentangling theorem leads to 
\begin{equation}
\langle\exp(\gamma a^{\dag}(t))\exp(-\gamma^*a(t))\rangle
\equiv\exp(-(2\kappa t-\frac{1}{2})\mid\gamma\mid^2)\langle\exp(\gamma a^{\dag}
-\gamma^* a)\rangle.
\end{equation}

Note that the Fourier transform of the left hand side yields the
quasi distribution known as the P-function of the system. Thus from (17) we conclude
that the P-function at time such that $2\kappa t =\frac{1}{2}$ is equal to the Wigner
function at $t=0$ and the P-function at time given by $2\kappa t=1$ is equal
to the Q-function at $t=0$. This implies that all {\it nonclassical} effects \cite{fifteen} will
{\it disappear} at times given by $2\kappa t\geq1$. Furthermore, the {\it
P-function} definitely {\it exists} as an {\it ordinary function} in the interval $1\geq 2\kappa
t\geq\frac{1}{2}$ though it can be negative. Eq.(16) also shows that the Wigner
function at time $t$ is equal to the s-parameterized distribution \cite{sixteen} $\Phi$s at
time $t=0$. This is because the s-parameterized distribution is the Fourier
transform of $\exp[s|\gamma |^2/2]\langle\exp(\gamma a^{\dagger}-\gamma^{*}a)\rangle$.
Clearly, for our problem, $s$ is equal to $-4\kappa t$. Note that for $s=-1$, we get
the Q-function, i.e. the Wigner function at time $2\kappa t=\frac{1}{2}$ is equal to
the Q-function at $t=0$. 

	We note in passing that if $\Gamma$ is related to $\kappa$ via the
relation

\begin{equation}
\frac\Gamma\kappa = \frac{\bar{n}}{(\bar{n}+1)}\leq 1,
\end{equation}
then the model (7) describes the interaction with a thermal bath
\cite{{seventeen},{eighteen}}.  However,  $\Gamma$ could exceed $\kappa$ as we are describing a pumped
environment. We could thus refer to the model (7) without the condition (18) as
the nonthermal and phase sensitive environment.

In summary, we have shown how the introduction of a gain medium can produce very
remarkable modifications in the dissipative dynamics of a superposition of
mesoscopic states. We demonstrated how to achieve classicality and
diagonalization in coherent state basis.

The author thanks R.P. Singh, S. Menon for the beautiful graphics and J. Kupsch,
W. Schleich for discussions on decoherence.


\newpage
\begin{figure}
\caption{Diagonalization in coherent state basis of a Cat State ~$(\mid\beta\rangle+\mid-\beta\rangle)$. 
	These frames show the behavior of the Wigner function as a function of $\alpha=x+iy$.
	The plots on the left show the results  [Eq.(13)]  for the new model of decoherence
	due to a controlled environment consisting of a gain medium,  whereas
	the plots on right show the results  [Eq.(5)]  for the standard model of
	decoherence.  The z-axis gives the numerical values of the Wigner function. The plot a0 gives the Wigner function at time t = 0. The
	subsequent plots are for increasing times for $\delta =2\kappa t = 0.1$
	for a1, b1;  0.5 for a2, b2;  4.0 for a3, b3. We have set  $\beta =3$ 
	for all the plots.   }	
\end{figure}	
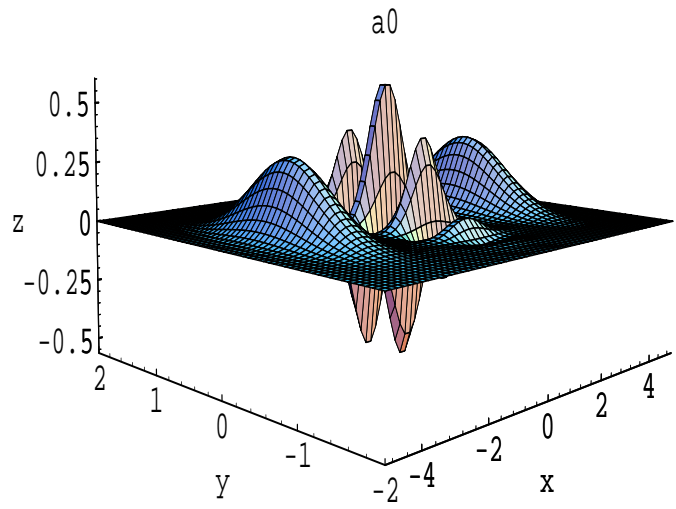
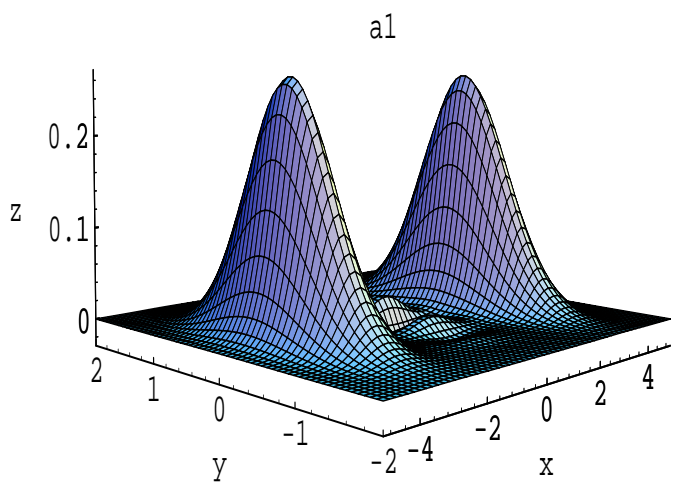
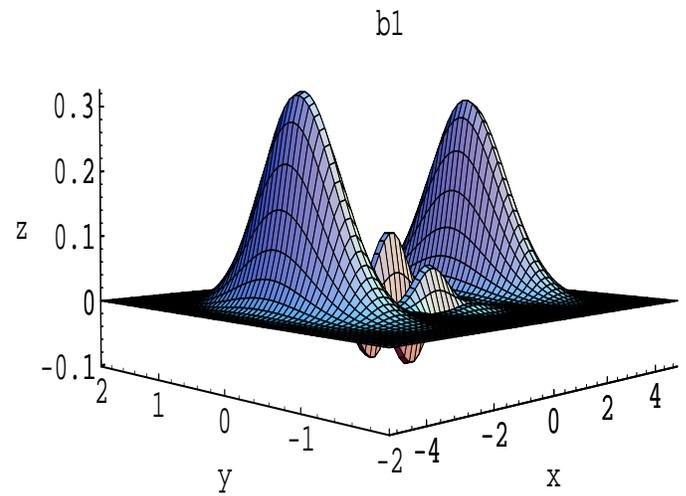
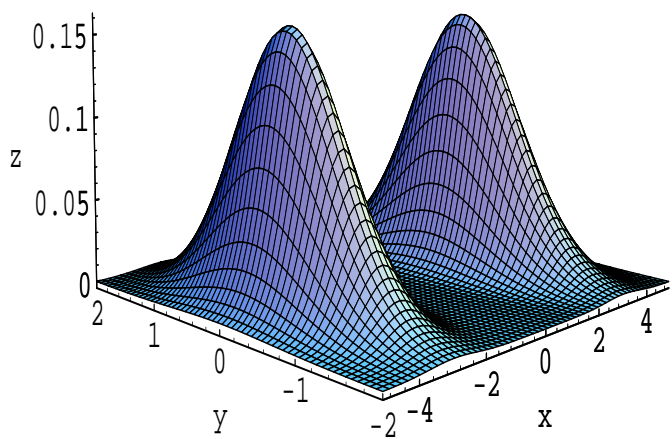
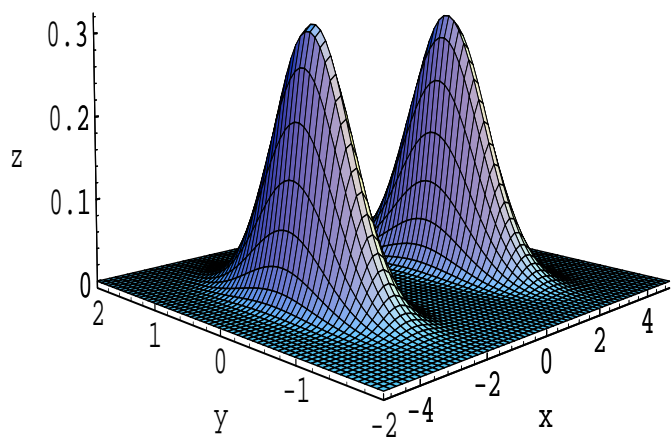
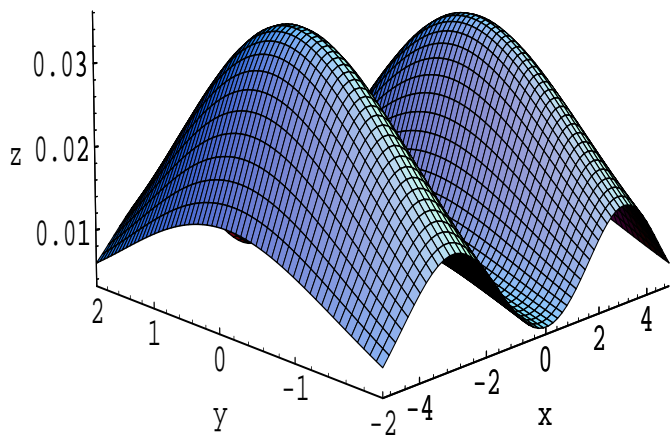
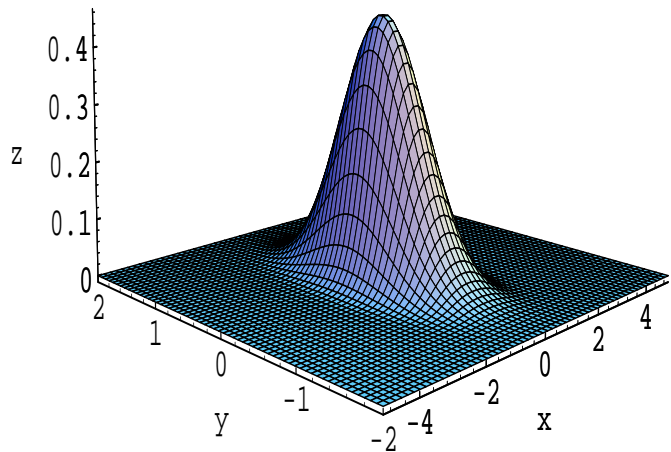
\end{document}